\documentclass[prl,twocolumn,showpacs,a4paper]{revtex4}
\usepackage{amssymb}
\usepackage{epsfig}
\usepackage{amsmath}
\usepackage{graphicx}
\usepackage{bm}
\usepackage{times}
\usepackage{color}

\begin{document}

\title{Optical Lattices with Micromechanical Mirrors}

\author{K. Hammerer}
\author{K. Stannigel}
\author{C. Genes}
\author{P. Zoller}
\affiliation{Institute for Theoretical Physics, University of Innsbruck, and Institute
for Quantum Optics and Quantum Information, Austrian Academy of Sciences,
Technikerstrasse 25, 6020 Innsbruck, Austria}
\author{P. Treutlein}\altaffiliation[Present address: ]{Department of Physics, University of Basel, Switzerland. E-mail: philipp.treutlein@unibas.ch}
%\email[E-mail: ]{treutlein@lmu.de}
\author{S. Camerer}
\author{D. Hunger}
\author{T. W. H{\"a}nsch}
\affiliation{Max-Planck-Institute of Quantum Optics and Faculty of Physics, Ludwig-Maximilians-Universit\"at M\"unchen, Schellingstr. 4, D-80799 Munich, Germany}

\newcommand{\sm}{{\scriptscriptstyle-}}
\renewcommand{\sp}{{\scriptscriptstyle+}}

\date{\today}

\begin{abstract}
We investigate a setup where a cloud of atoms is trapped in an optical lattice potential of a standing wave laser field which is created by retro-reflection on a micro-membrane. The membrane vibrations itself realize a quantum mechanical degree of freedom. We show that the center of mass mode of atoms can be coupled to the vibrational mode of the membrane in free space, and predict a significant sympathetic cooling effect of the membrane when atoms are laser cooled. The controllability of the dissipation rate of the atomic motion gives a considerable advantage over typical optomechanical systems enclosed in optical cavities, in that it allows a segregation between the cooling and coherent dynamics regimes. The membrane can thereby be kept in a cryogenic environment, and the atoms at a distance in a vacuum chamber.
\end{abstract}

\maketitle

\begin{figure}[t]
\begin{center}
\includegraphics[width=1\columnwidth]{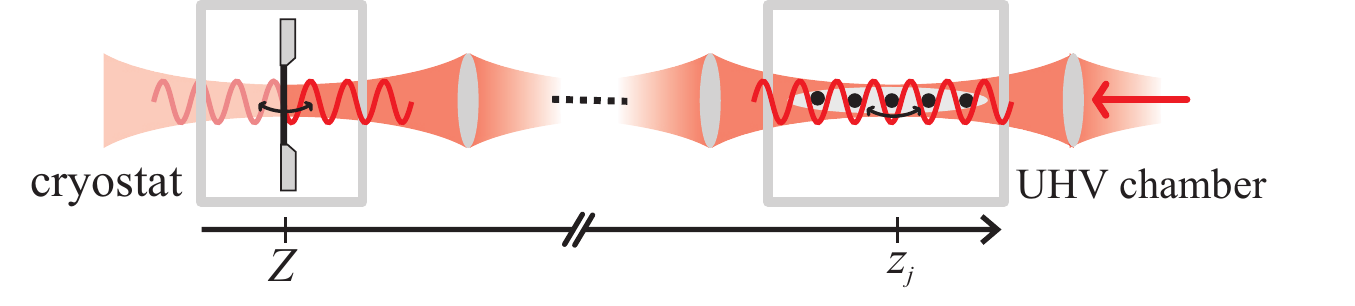}
\end{center}
\caption{A laser field impinging from the right is partially reflected off a dielectric membrane and forms a standing wave optical potential for an atomic ensemble. Vibrations of the membrane's fundamental mode will shift the standing wave field, shaking atoms in the optical lattice. Conversely, oscillations of the  atomic cloud (center of mass motion) will change the intensity of left/right propagating field components, thus shaking the membrane via changing the radiation pressure on it. The membrane can be kept in a cryogenic environment, and the atoms at a distance in a vacuum chamber.}
\label{Fig:setup}
\end{figure}

Optical lattices for ultracold atoms \cite{Bloch2005} are generated by standing laser light waves, obtained e.g.\ by retroreflecting a running wave laser beam from a mirror. The intensity variation in the standing wave gives rise to a position dependent AC Stark shift, thus providing a periodic array of optical microtraps for the atomic motion. In the description of atomic dynamics in far-off-resonant lattices, e.g.\ in deriving the Hubbard dynamics of bosonic or fermionic atoms \cite{Bloch2005}, the optical potentials are typically represented as c-number fields. The backaction of the atomic motion on the laser fields and spontaneous emission into vacuum modes appear as weak decoherence mechanisms, in particular heating of atoms.
Thermal and other vibrations of the mirror reflecting the laser light provide an additional source of dissipation. % and imperfections.
%In a lattice with moderate detuning, on the other hand, the back-action of atomic motion on the light fields is substantial \cite{Raithel1998}.
%Moreover, spontaneous emission can be harnessed for laser cooling of the atoms to the motional ground state \cite{Kerman2000}.
%In light of recent experimental progress with micromechanical oscillators with attached mirrors at low temperatures
In light of recent experimental progress in the field of optomechanics \cite{Kippenberg2008,Marquardt2009}, a new generation of optical lattice experiments is in sight, where the standing light waves are generated by reflection from a {\em micromechanical mirror} or reflecting {\em dielectric membrane} \cite{Thompson2008}, whose motion must be described {\em quantum mechanically} at low temperature, and which is coupled via the moving lattice to a {\em distant} ensemble of cold atoms.
In a lattice with moderate detuning, the back-action of atomic motion on the light fields can be substantial \cite{Raithel1998}. This results in a sizable coupling of atomic motion and mirror vibrations, mediated by the quantum fluctuations of the lattice laser light.
Below we develop a fully quantum theoretical description of such a
%{\em half cavity setup}
setup (cf.\ Fig.~\ref{Fig:setup}). %, where the atomic motion in the lattice is coupled to the quantum dynamics of the mirror, mediated by the quantum fluctuations of the lattice laser light.
In particular, we derive a Quantum Stochastic Schr\"odinger Equation (QSSE) and thus a Master Equation for a setup where a {\em coherent} coupling of the mirror to the collective center-of-mass motion of atoms in the lattice emerges, and where the mirror can be {\em sympathetically cooled} to the motional ground state by laser cooling the atoms.
We envision that the proposed system presents only one instance where the well-developed tools for laser cooling of atoms are used to provide a mechanism for sympathetic cooling of polarizable particles (here the dielectric membrane).

%\newpage

\begin{figure}[t]
\begin{center}
\includegraphics[width=1\columnwidth]{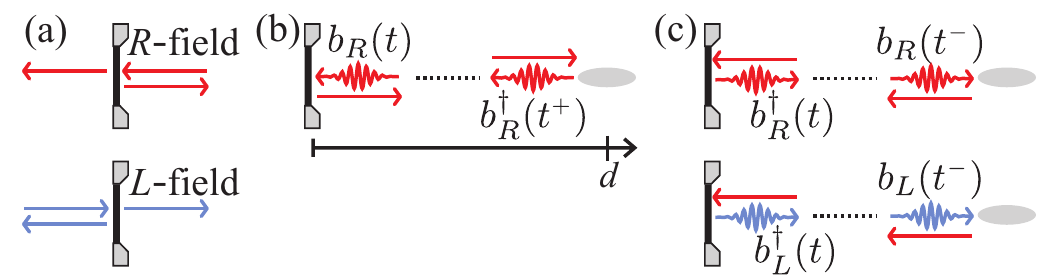}
\end{center}
\caption{(a) $R$- and $L$-modes for fields impinging from the right and left, respectively. All but the laser driven $(\omega_l,R)$-mode are initially in vacuum. (b) Mediated action of atomic COM fluctuations on the membrane: At the advanced time $t^\sp$ atoms absorb a photon from the right propagating component of the laser field (with amplitude $\sqrt{\mathfrak{r}}\alpha$, straight arrow) and reemit a sideband photon to the $R$-field ($b^\dagger_R(t^\sp)$, wavy arrow), receiving a momentum kick of $2\hbar k_l$, cf.\ the term in Eq.~\eqref{Eq:QSSEtplus}. At a later time $t$ the sideband photon is annihilated ($b_R(t)$) and transferred back to the laser field giving a momentum kick to the membrane, cf.\ first term in Eq.~\eqref{Eq:QSSEt}. (c) Reverse action: Via corresponding emission and reabsorption processes as given in Eqs.~\eqref{Eq:QSSEt} and \eqref{Eq:QSSEtminus} membrane vibrations affect atomic COM fluctuations. In these processes both the $R$- (upper panel) and the $L$-field (lower panel) contribute, effectively mediating a stronger action than the process shown in (b).
 %Schematic of the setup: Both, membrane and atoms couple to the $R$- and $L$-component  of the EM fields impinging from the right and left, indicated here by red (solid) and grey (dashed) arrows respectively. Vacuum fluctuation `noiselets' in the $R$-component couple to atomic position fluctuations at time $t^\sm=t-d/c$. At time $t$ $R$- and $L$-noiselets couple to membrane fluctuations, and finally at $t^\sp=t+d/c$ also with atoms. Emission and reabsorption of sideband photons gives rise to an effective coupling between membrane and atomic COM motion, along with radiation pressure induced decoherence and certain effects which are characteristic for (partially) unidirectionally coupled, cascaded quantum systems.
 }
\label{Fig:Schematic}
\end{figure}

\paragraph*{Model:} Consider the setup shown in Fig.~\ref{Fig:setup}a. A laser (frequency $\omega_l$) is shined on and partially retro-reflected from a dielectric membrane to form a standing wave that acts as a one-dimensional optical lattice trap for a cloud of atoms. The Hamiltonian for the system is $H_{\rm f\!ull}=H+H_{3D}$, where $H$ contains in a one-dimensional treatment the coupling of atoms and membrane vibrations to the paraxial field along the $z$ direction as shown in Fig.~\ref{Fig:setup},
%\begin{multline*}
\begin{align}\label{Eq:Ham}
&H=\hbar\omega_m a_m^\dagger a_m+
\sum_{j=1}^N{\textstyle\frac{p_j^2}{2m_{at}}}-\sum_{j=1}^N{\textstyle\frac{\mu^2}{\hbar\delta}}E^-(z_j,t)E^+(z_j,t)\nonumber\\
&+\epsilon_0A(n^2-1)\left[E^-(\textstyle{\frac{l}{2}})E^+(\textstyle{\frac{l}{2}})
-E^-(-\textstyle{\frac{l}{2}})E^+(-\textstyle{\frac{l}{2}})\right]Z,
\end{align}
%\end{multline*}
while coupling to all other field modes is subsumed in $H_{3D}$. The first two terms in Eq.~\eqref{Eq:Ham} represent the energy of the fundamental vibrational mode of the membrane with frequency $\omega_m$ and annihilation operator $a_m$, and the kinetic energy of atoms with momentum $p_j$ and mass $m_{at}$ . The third term is the level shift operator \cite{Cohen1992} describing the dispersive interaction of the $N$ two level atoms (with dipole matrix element $\mu$) with off-resonant field modes with detuning $\delta$ from the atomic resonance within a frequency band $\vartheta\ll\delta$ \cite{Gardiner2000}. In this treatment electronically excited states are already adiabatically eliminated. The second line in Eq.~\eqref{Eq:Ham} is the potential of the radiation pressure force on a slab of thickness $l$, cross section $A$, index of refraction $n$, and a small excursion $Z$ of the membrane from the equilibrium. The radiation pressure force is proportional to the difference of the intensities on either side (at $\pm\frac{l}{2}$) of the membrane. The electric field is $E(z)=E_R(z)+E_L(z)$ and consists of the two 1D continua of plane wave modes impinging from the right $(R)$ and left $(L)$, respectively, cf.\ Fig.~\ref{Fig:Schematic}a. The explicit mode functions can be found in \cite{Vogel2006}, and will be denoted by $A_\alpha(k,z)\,(\alpha=L,R)$ consistent with the notation used there. Thus, $E_\alpha^+(z,t)=\mathcal{E}\int\!\!d\omega A_\alpha(k,z) b_{\omega,\alpha}e^{-i\omega t}$ with $[b_{\omega,\alpha},b^\dagger_{\bar\omega,\beta}]=\delta_{\alpha,\beta}\delta(\omega-\bar\omega)$ and $k=\omega/c$. We are using a narrow band approximation $\mathcal{E}=\sqrt{\frac{\hbar \omega_l}{4\pi c\epsilon_0 A}}$
%%
%% KAI: PT hat im Nenner \pi, du hattest 4\pi. Check!
%%
and take the fields in an interaction picture with respect to the free field Hamiltonian.
The contribution of other than paraxial modes contained in $H_{3D}$ can be largely suppressed in the following. Its effect will be relevant and addressed separately in the context of momentum diffusion of atoms in the lattice. The driving laser of frequency $\omega_l$ is at this point included by assuming the field to be in a coherent state $|\alpha\rangle=D|\mathrm{vac}\rangle$ (with $D=\exp(\alpha^* b_{\omega_l,R}-\mathrm{h.c.})$), corresponding to a laser field of power $P=\hbar\omega_l\alpha^2/2\pi$ (with photon flux $\alpha^2/2\pi$) impinging from the right.

\paragraph*{Quantum Stochastic Schr\"odinger Equation:} We are interested in treating those processes only whose amplitudes are enhanced by the laser amplitude $\alpha$ in either first or second order. In order to identify those we apply a Mollow transformation \cite{Mollow1975} and move to a displaced picture, where the field is effectively in vacuum  $|\mathrm{vac}\rangle=D^\dagger|\alpha\rangle$ and the new Hamiltonian is accordingly $H'=D^\dagger HD$. In $H'$ the second order terms in $\alpha$ provide the lattice potential (with trap depth $V_0=\frac{4 \mu^2 \mathcal{E}^2\alpha^2\sqrt\mathfrak{r}}{\hbar\delta}$
and trap frequency $\omega_{at}=\sqrt{\frac{2V_0k_l^2}{m_{at}}}$ for a membrane power reflectivity $\mathfrak{r}$), and an unimportant mean radiation pressure force on the membrane. The terms of first order in $\alpha$, which contain the coupling of the motion of membrane and atoms to electromagnetic (EM) field modes initially in vacuum, are treated in a Lamb-Dicke expansion in $(k_l\ell_{at})$ where $\ell_{at}=\sqrt{\frac{\hbar}{m_{at}\omega_{at}}}$. The result can be interpreted as a (Stratonovich) quantum stochastic Schr\"odinger equation (QSSE) \cite{Gardiner2000} with \textit{time delays}
%\begin{multline}
\begin{subequations}\label{Eq:QSSE}
\begin{align}
  &\textstyle{\frac{d}{dt}}|\Psi\rangle=-iH(t,t^\sm,t^\sp)|\Psi\rangle\nonumber\\
  &=\Big\{-iH_{sys}
  +x_{at}\big(\sqrt{g_{at,R}}\,b_R(t^\sp)-\mathrm{h.c.}\big)\label{Eq:QSSEtplus}\\
  &+x_m \big(i\sqrt{g_{m,R}}\,b_R(t)-\sqrt{g_{m,L}}\,b_L(t)-\mathrm{h.c.}\big)\label{Eq:QSSEt}\\
  &-x_{at} \big(\sqrt{g_{at,R}}\,b_R(t^\sm)+i\sqrt{g_{at,L}}\,b_L(t^\sm)-\mathrm{h.c.}\big)
  \Big\}|\Psi\rangle,\label{Eq:QSSEtminus}
\end{align}
\end{subequations}
where $H_{sys}=\omega_{m}a_m^\dagger a_m +\omega_{at}a_{at}^\dagger a_{at}$ denotes the system Hamiltonian for the free evolution of the membrane and the atomic center of mass (COM) mode, with annihilation operator $a_{at}$ for the later. The remaining terms denote the coupling of position fluctuations $x_{m(at)}=(a_{m(at)}+\mathrm{h.c.})/\sqrt{2}$ of these modes to vacuum fluctuations of the EM field described by slowly varying field operators $b_{\alpha}(t)=\int \frac{d\omega}{\sqrt{2\pi}} b_\alpha(\omega)\exp[-i(\omega-\omega_l)t]$ $(\alpha=R,L)$, evaluated at time $t$ as well as at \textit{advanced} and \textit{retarded} times $t^{\scriptstyle{\pm}}=t\pm \frac{d}{c}$, where $d$ is the distance between the atomic ensemble and the mirror, cf.\ Fig.~\ref{Fig:Schematic}b. In this Hamiltonian the linear extension of the ensemble is neglected and we used the approximation $b(t\pm \frac{z_j}{c})\simeq b(t\pm \frac{d}{c})$ for all atoms \footnote{This is well justified if the `sideband wavelength' $c/\omega_{m(at)}$ is smaller than the linear extension of the atomic ensemble.}. The coupling constants of membrane vibrations and atomic COM position fluctuations to the EM field are given by
\begin{align*}
  g_{m,R}&=2(\alpha k_l\ell_m)^2\mathfrak{r}^2/\pi, & g_{at,R}=2\pi N(\omega_{at}/4\alpha k_l\ell_{at})^2,
\end{align*}
where $\ell_m=\sqrt{\frac{\hbar}{m_m\omega_m}}$ ($m_m$ is the effective mass of the membrane), and $g_{m(at),L}=\frac{\mathfrak{t}}{\mathfrak{r}}\, g_{m(at),R}$ with the membrane transmittivity $\mathfrak{t}=1-\mathfrak{r}$. In Fig.~\ref{Fig:Schematic}b and c we provide an intuitive interpretation of each term in the QSSE~\eqref{Eq:QSSE} as an (Anti)Stokes-backscattering process, where a laser photon is absorbed or emitted together with, respectively, an emission or absorption of a sideband photon at frequency $\omega_l\pm\omega_{m(at)}$ in one of the initially empty modes, along with a momentum transfer of $2\hbar k_l$ to either the membrane or the atomic COM mode. The effective mediated coupling between atoms and membrane, as well as radiation pressure noise on each system, will eventually result from second order processes, where sideband photons are emitted and reabsorbed, Fig.~\ref{Fig:Schematic}b and c.

\paragraph*{Cascaded System Master Equation} The corresponding effects of these second order processes can be seen explicitly when Eq.~\eqref{Eq:QSSE} is transformed into an equivalent master equation (MEQ) \textit{without} time delays in the limit $t^{\scriptstyle{\pm}}\rightarrow t$, and under the Born-Markov approximation for the EM field in vacuum. The technicalities of this procedure, which requires some care regarding correct time ordering, are outlined below. The result is a master equation for the density operator $\rho$ of the membrane and the COM motion of the form
\begin{align}\label{Eq:MEQ}
\dot{\rho}=-i[H_{sys}+gx_{at}x_m,\rho]+C\rho+L_m\rho+L_{at}\rho.
\end{align}
This MEQ contains, firstly, an \textit{effective atom-membrane interaction} with a coupling $g=2(\sqrt{g_{{m},R}g_{{at},R}}+\sqrt{g_{{m},L}g_{{at},L}})=\omega_{at}\sqrt{\frac{Nm_{at}}{m_{m}}}$.
%%
%% CHECK factor of 1/2
%%
  %($g=\omega_{at}\sqrt{Nm_{at}/m_{m}}$)
Here we assume comparable oscillation frequencies $\omega_m\simeq\omega_{at}$. While the ratio of masses $\sqrt{\frac{m_{at}}{m_m}}$ is always small under reasonable conditions, the rate of coherent coupling $g$ can still be significant thanks to the collective enhancement by $\sqrt{N}$.
%E.g. for a membrane with an effective mass of $m_m=0.4$~ng and resonance frequency of $\omega_m/2\pi=1.3$~MHz as in \cite{Thompson2008}, and a large number of $N=10^7$ $^{87}$Rb atoms trapped in a deep lattice with $\omega_{at}\simeq\omega_m$ a coupling of $g=10$~kHz can be achieved.

Secondly, we find a contribution
\[
C\rho=\textstyle{\frac{i\mathfrak{t}g}{2}}\big([x_m,x_{at}\rho]-[\rho x_{at},x_m]\big),
\]
which is proportional to the membrane transmittivity $\mathfrak{t}$. The peculiar form of this term is in fact generic for \textit{cascaded} quantum systems \cite{Gardiner2000,Gardiner1993}. Its effect becomes clear e.g.\ when looking at the evolution of mean values according to Eq.~\eqref{Eq:MEQ}, $\langle\dot p_{at}\rangle=g\langle x_{m}\rangle+\cdots$ while $\langle\dot p_{m}\rangle=g\mathfrak{r}\langle x_{at}\rangle+\cdots$, where dots stand for contributions of $H_{sys},L_m$ and $L_{at}$. Thus, the action of atoms on the membrane is reduced by a factor $\mathfrak{r}$ as compared to the action of membrane fluctuations on atoms \footnote{Note the coupling strength $g$ itself tends to zero for $\mathfrak{r}\rightarrow 0$, as the lattice depth and therefore also $\omega_{at}$ vanish in this limit and $g\propto\omega_{at}$.}. This can be understood from the fact that action of membrane on atoms is mediated through both fields, $R$ and $L$, while only the $R$ field contributes to the action of atoms on the membrane, see Fig.~\ref{Fig:Schematic}.

In the limiting case $\mathfrak{r}=1$, that is for an ideal mirror, the term $C\rho$ does not contribute and the interaction is only due to the Hamiltonian part of Eq.~\eqref{Eq:MEQ}. The scaling of $g$ as given above can then be understood in terms of a simple consideration in terms of quasi-static field modes: The mirror displacement $Z(t)$ provides a time-dependent boundary condition for the EM field, such that the standing wave laser field is $E(z)\propto\sin[k_l(z-Z)]$. The corresponding lattice potential for the atoms is $V(z_i)=\sum_iV_0\sin[k_l(z_i-Z)]^2\simeq \sum_i V_0k_l^2(\delta z_i^2+2\delta z_i\delta Z)$, where $\delta z_i$ and $\delta Z$ denote small fluctuations of atoms and mirror around their respective equilibrium positions. This is on the one hand just the harmonic potential for the atoms with a trap frequency $\omega_{at}=(2V_0 k_l^2/m_{at})^{1/2}$ (the corresponding term for the mirror can be absorbed into $\omega_m$), and on the other hand a {\it direct} coupling of the mirror and the COM motion of the atoms with a strength $2V_0k_l^2\sum_i\delta z_i\delta Z=\hbar gx_{at}x_{m}$, correctly reproducing $g=\sqrt{N}\,2V_0k_l^2\ell_{at}\ell_{m}/\hbar=\omega_{at}\sqrt{\frac{Nm_{at}}{m_m}}$. However, for the realistic case of a membrane with reflectivity $\mathfrak{r}<1$ such considerations can hardly lead to the correct master equation \eqref{Eq:MEQ} \footnote{For $\mathfrak{r}<1$ a more careful reasoning in terms of quasi-static field modes can in fact reproduce the correct equations of motion for mean values.}.

The strength of coherent coupling has to be compared to various rates of decoherence, described in Eq.~\eqref{Eq:MEQ} by a number of Lindblad terms $L_{m},\,L_{at}$. One channel of decoherence is caused by coupling to the EM vacuum field, which is contained in the QSSE in Eq.~\eqref{Eq:QSSE} and results in radiation pressure induced momentum diffusion described by Lindblad terms $L^{dif\!f}_{at(m)}\rho=\frac{1}{2}\gamma^{dif\!f}_{at(m)}D[x_{at(m)}]\rho$, where $D[a]\rho=2a\rho a^\dagger-a^\dagger a\rho-\rho a^\dagger a$. The atomic momentum diffusion rate is $\gamma_{at}^{dif\!f}=(k_l\ell_{at})^2\gamma_{se}V_0/\hbar\delta$ with a natural linewidth $\gamma_{se}$ and a detuning $\delta$ from atomic resonance. Note that a momentum diffusion rate $\gamma_{at}^{dif\!f}$ for the atomic COM motion is equal to the well-known single atom diffusion rate \cite{Ashkin2006} only if all interactions between atoms are neglected. In the 1D treatment of the QSSE~\eqref{Eq:QSSE} these interactions are in fact overestimated, as they do not fall off with the distance between atoms. Only a more careful 3D treatment, taking into account all terms neglected in $H_{3D}$, reveals the correct distance dependence of dipole-dipole interactions and allows to ignore these effects for a dilute sample. The diffusion rate of the membrane mode due to radiation pressure is $\gamma^{dif\!f}_m=(4 \mathfrak{r}P/m_mc^{2})(\omega_l/\omega_m)$ (where $P$ is the laser power). This last result is consistent with calculations in \cite{Karrai2008} on the effects of radiation pressure on a mirror in free space.

On top of these decoherence channels there will be thermal heating of the membrane due to clamping losses and laser absorption, which drives the vibrational mode under consideration to thermal equilibrium at temperature $T$, i.e. $L^{th}_m\rho=\tfrac{\gamma_m}{2}(\bar{n}+1)D[a_m]\rho+\tfrac{\gamma_m}{2}\bar{n}D[a^\dagger_m]\rho$, with thermal occupation $\bar{n}=k_BT/\hbar\omega_m$, and $\gamma_m=\omega_m/Q_m$ for a mechanical quality factor $Q_m$. For atoms in turn, it is important to note that the dissipation processes $L_{at}$ can to a large extent be \textit{engineered}. In particular, it is possible to use Raman sideband \textit{laser cooling} to cool atoms individually close to their ground states in the optical lattice potential \cite{Kerman2000}. In the master equation we account for this by adding a Lindblad term for the COM mode, $L^{cool}_{at}\rho=\tfrac{1}{2}\gamma^{cool}_{at}D[a_{at}]\rho$ with a Raman-cooling rate $\gamma^{cool}_{at}$. If laser cooling is switched off, only the background decoherence at a rate $\gamma^{dif\!f}_{at}$ remains.
The dynamics of the master equation~\eqref{Eq:MEQ} thus provides two interesting regimes: with efficient laser cooling of atoms it is possible to \textit{sympathetically cool} the mirror, while for $\gamma^{diff}_{at}\ll g$ a regime of \textit{coherent evolution} becomes accessible, limited only by the decoherence rate of the membrane $\gamma_m^{th}=\gamma_m \bar{n}$.

\paragraph*{Case Study} We estimate the magnitude of the relevant processes as follows: For a small SiN membrane of dimensions 150$\,\mu$m\,$\times$\,150\,$\mu$m\,$\times$\,50\,nm exhibiting a mechanical resonance at $\omega_m=2\pi\times0.86$\,MHz with an effective mass of $m_m=8\times10^{-13}$\,kg and a high mechanical quality of $Q_m=10^7$ at $T\lesssim2$\,K, a power reflectivity $\mathfrak{r}= 0.31$ can be achieved for the wavelength of $\lambda=780$\,nm relevant for $^{87}$Rb \cite{Thompson2008}. A lattice beam with power $P=7$\,mW and a waist of 230\,$\mu$m at the position of atoms, detuned by $\delta=2\pi\times1$\,GHz, can provide a sufficient lattice depth such that the longitudinal trap frequency $\omega_{at}\simeq\omega_m$. For an ensemble of $N= 3\times 10^8$ atoms \cite{Kerman2000} a strength of coherent coupling $g=40$\,kHz can be achieved. For the decoherence due to radiation pressure noise one finds a momentum diffusion at rate $\gamma_m^{dif\!f}=52$\,Hz for the membrane, and an atomic momentum diffusion in the lattice at rate $\gamma_{at}^{dif\!f}=16$\,kHz. For the membrane the dominating dissipative effect will clearly be thermal decoherence at rate $\gamma_m^{th}=4$\,MHz (24\,kHz) at room temperature (2\,K). On the other hand, Raman sideband cooling of atoms at a fast rate $\gamma_{at}^{cool}\simeq 20$\,kHz is possible  \cite{Kerman2000}. For these parameters, a regime where $\omega_m=\omega_{at}\gg g\gtrsim\gamma_{at}^{cool} \simeq \gamma_m^{th}   \gtrsim \gamma_{at}^{diff} \gg \gamma^{diff}_{m}$ is accessible.
To suppress trap loss due to light-assisted collisions, blue detuning of the lattice laser is advantageous \cite{Kerman2000}. Transversally, the atoms can be confined in a far-detuned 2D lattice, so that atomic densities of order $10^{12}$~atoms$/\textrm{cm}^3$ are realistic \cite{DePue1999}. A concern is the spread of vibrational frequencies $\Delta \omega_{at}\simeq 2\pi\times 24$\,kHz across the ensemble due to the Gaussian intensity profile of the laser. It can be reduced by using a beam with a flat top profile.
Due to its modularity, the presented setup is very flexible, e.g.\ several atomic ensembles could be trapped in multiple foci of the same lattice laser, thereby enhancing $N$ and thus $g$.

\paragraph*{Sympathetic Membrane Cooling} In order to evaluate the corresponding efficiency of sympathetic cooling of the membrane via laser cooling of atoms, we solve the master equation \eqref{Eq:MEQ} for the steady state occupation $\bar{n}_{ss}=\langle a_m^\dagger a_m\rangle_{ss}$. For the parameters given above one finds a cooling factor $f=\bar{n}/\bar{n}_{ss}\simeq 2\times10^4$, which is sufficient for \textit{ground state cooling} ($\bar{n}_{ss}\simeq 0.8$) of the membrane when starting from 500\,mK. In order to get some more insight into the cooling efficiency we consider the weak-coupling limit $\gamma_{at}^{cool}\gg g$. In this case we can eliminate the atomic COM mode adiabatically along the lines of the treatments of the equivalent problem of optomechanical laser cooling \cite{Kippenberg2008,Marquardt2009}. For the mean occupation one finds a rate equation $\frac{d}{dt}\langle a^\dagger_ma_m\rangle=-\Gamma_m(\langle a^\dagger_ma_m\rangle-\bar{n}_{ss})$ with an effective cooling rate $\Gamma_m=\gamma_m+\frac{\mathfrak{r}g^2}{2\gamma_{at}^{cool}}$ and a final occupation
$
\bar{n}_{ss}\simeq\frac{\gamma_m}{\Gamma_m}\bar{n}+\big(\frac{\gamma_{at}^{cool}}{4\omega_m}\big)^2.
$
For large enough cooling rate the thermal contribution can be suppressed and the limitation is only due to Stokes-scattering processes. As in laser cooling of ions or optomechanics, this can be suppressed in the resolved sideband limit, which amounts here to having $\gamma_{at}^{cool}\ll\omega_m$. Under this condition ground state cooling is possible.

If laser cooling of atoms is switched off $(\gamma_{at}^{cool}=0)$, a regime of coherent coupling is accessible, at least for cryogenic temperatures, where $g\gtrsim \gamma_m^{th},\gamma^{diff}_{at}$. As compared to the usual optomechanical setup, where the equivalent parameter to $\gamma_{at}^{cool}$ is the cavity decay rate which is a fixed parameter, this is a qualitatively new feature of the setup considered here. It is well known, and has been extensively studied in various contexts \cite{Hammerer2009}, that the given coupling Hamiltonian $\sim x_{at}x_m$ -- which becomes $\sim(a_ma_{at}^\dagger+a_{m}^\dagger a_{at})$ in the rotating wave approximation ($\omega_m\gg g$) -- allows for a coherent state exchange of the two systems.
\begin{figure}[t]
\begin{center}
\includegraphics[width=0.75\columnwidth]{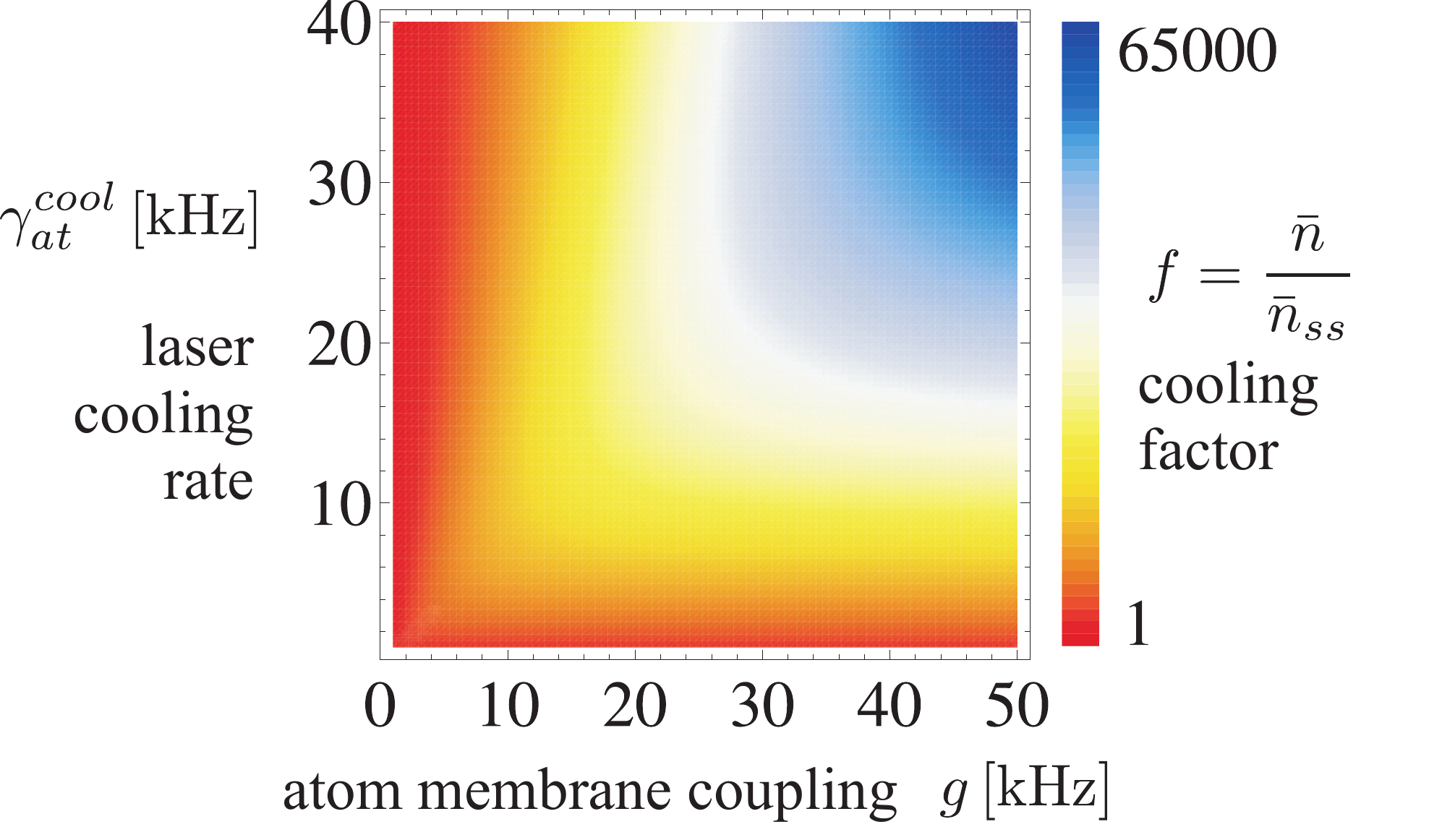}
\end{center}
\caption{Cooling factor $f=\bar{n}/\bar{n}_{ss}$ versus effective coupling $g$ between the atomic COM motion and the membrane vibrations and rate of Raman sideband laser cooling $\gamma_{at}^{cool}$, as determined from an exact numerical solution of the effective master equation~\eqref{Eq:MEQ} for a resonant system $\omega_m=\omega_{at}$ and a mode of the membrane with $\omega_m=2\pi\times 0.86$\,MHz and $Q_m=10^7$ at $T=2K$. The calculation includes radiation pressure induced momentum diffusion at rate $\gamma^{dif\!f}_{at}=16$\,kHz for atoms and $\gamma^{dif\!f}_{m}=52$\,Hz for the membrane. The parameters $g\simeq40$\,kHz and $\gamma_{at}^{cool}\simeq20$\,kHz discussed in the text provide a cooling factor of $10^4$, and yield a steady state occupation $\bar{n}_{ss}$ below one if the system is precooled to below 500\,mK. At room temperature the sympathetic cooling effect will be still clearly observable.}
\label{Fig:cooling}
\end{figure}

\paragraph*{Appendix: Conversion of the QSSE from Stratonovich to Ito form} We still need to explain how to transform the QSSE with time delays to the Markovian master equation \eqref{Eq:MEQ}. We first transform the Stratonovich QSSE to an Ito QSSE by first integrating Eq.~\eqref{Eq:QSSE} up to a time $t+\Delta t$ (such that $\Delta t\gg\tau\gg1/\vartheta$, where $\vartheta$ is the bandwidth of modes) and then expand to second order,
\begin{align}
  &U(\Delta t)|\Psi(t)\rangle={\textstyle \left\{1-i{\int_t^{t+\Delta t}}%\hspace{-3mm}
  dt_1H(t_1,t_1^\sm,t_1^\sp)\right.}\label{Eq:Ito}\\
  &\quad{\textstyle\left.-\int_t^{t+\Delta t}%\hspace{-4mm}
  dt_1\int_t^{t_1}%\hspace{-2mm}
  dt_2H(t_1,t_1^\sm,t_1^\sp)H(t_2,t_2^\sm,t_2^\sp)\right\}|\Psi(t)\rangle}\nonumber\\
  %&=\Big\{\mathds{1}-iH_{sys}\Delta t-i\sqrt{\Gamma_m}x_m\Delta B^\dagger(t)\nonumber\\
  &=\Big\{1-i\left[H_{sys}+(2 \sqrt{g_{{m},R}g_{{at},R}}+\sqrt{g_{{m},L}g_{{at},L}})x_mx_{at}\right]\Delta t\nonumber\\
  %%
  %% KAI: wir wollten die g's doch so definieren, dass wir keine 2 bei 2\sqrt{g_{{m}... haben?
  %%
  &\quad+i\sqrt{g_{{m},R}}x_m\Delta B^\dagger_R +\left(\sqrt{g_{{m},L}}x_m-i\sqrt{g_{{at},L}}x_{at}\right)\Delta B^\dagger_L\nonumber\\
  &\quad-\textstyle{\frac{1}{2}}(g_{{m},R}+g_{{m},L})x_m^2\Delta t
  -\textstyle{\frac{1}{2}}g_{{at},L}x_{at}^2\Delta t
  \Big\}|\Psi(t)\rangle.\label{Eq:ItoQSSE}
\end{align}
We use here the Ito increments $\Delta B_\alpha(t)=\int_t^{t+\Delta t}dt' b_\alpha(t')$ and their property $\Delta B_\alpha(t)|\Psi\rangle\equiv 0$ for all times. In the double time integral care needs to be taken in dealing with the time delays. It is easy to check that $\int_t^{t+\Delta t}\hspace{-0mm}dt_1\int_t^{t_1}\hspace{-0mm}dt_2 b_\alpha(t_1)b_\beta^\dagger(t_2+\delta)=\theta(\delta)\delta_{\alpha,\beta}\Delta t$ \footnote{$\theta(\delta)=0,\,\frac{1}{2},\,1$ for $\delta\lesseqqgtr0$ is the step function.}. \textit{After} applying this rule it is possible to take the delay to zero, $t^{\scriptscriptstyle\pm}\rightarrow t$, as has been done in the second step of Eq.~\eqref{Eq:Ito}. The last equation~\eqref{Eq:ItoQSSE} can then be interpreted as an Ito QSSE, which can in turn be converted to the master equation~\eqref{Eq:MEQ} following standard procedures \cite{Gardiner2000}. Note that this procedure results in a seemingly {\em collectively enhanced} atomic momentum diffusion, which is an artefact of the 1D model adopted in Eq.~\eqref{Eq:Ham}. A more careful treatment based on a 3D model for the coupling of atoms to the EM field correctly yields the well known individual momentum diffusion of atoms in an optical lattice at the rate given in Eq.~\eqref{Eq:MEQ}, along with dipolar interactions which are suppressed at low densities \footnote{A 1D treatment misses the fall-off of inter-atomic interactions with the cubed distance, giving rise to an entirely collective diffusion.}.


\begin{thebibliography}{10}

\bibitem{Bloch2005} I. Bloch, Nature Phys. \textbf{1}, 23 (2005).



\bibitem{Kippenberg2008} T.J. Kippenberg and K. Vahala, Science \textbf{321}, 1172 (2008).

\bibitem{Marquardt2009} F. Marquardt and S.M. Girvin, Physics \textbf{2}, 40 (2009).

\bibitem{Thompson2008} J.D. Thompson et al., Nature \textbf{452} 72 (2008), A.M. Jayich et al., New J. Phys. \textbf{10}, 095008 (2008).

\bibitem{Raithel1998} G. Raithel et al., Phys. Rev. Lett. \textbf{81}, 3615 (2000).

\bibitem{Cohen1992} C. Cohen-Tannoudji, J. Dupont-Roc, G. Grynberg, \textit{Atom-Photon Interactions}, John Wiley \& Sons, 1992

\bibitem{Gardiner2000} C.W. Gardiner, P. Zoller, \textit{Quantum Noise}, Springer, 2000

\bibitem{Ashkin2006} A. Ashkin, \textit{Optical trapping and manipulation of neutral particles using lasers}, World Scientific, 2006

%\bibitem{Knoell1986} L. Kn\"oll et al., Opt. Soc. Am. B \textbf{3}, 1315 (1986), L. Kn\"oll et al., \textbf{42}, 503 (1990), M. Ley and R. Loudon, J. Mod. Opt. \textbf{34}, 227 (1987)

\bibitem{Vogel2006} W. Vogel, D.-G. Welsch, \textit{Quantum Optics}, Wiley, 2006


\bibitem{Mollow1975} B.R. Mollow, Phys. Rev. A \textbf{12}, 1919 (1975)

\bibitem{Kerman2000} A. Kerman et al., Phys. Rev. Lett. \textbf{84}, 439 (2000).

\bibitem{DePue1999} M.T. DePue et al., Phys. Rev. Lett. \textbf{82}, 2262 (1999).

\bibitem{Karrai2008} K. Karrai et al., Phys. Rev. Lett. \textbf{100}, 240801 (2008).

\bibitem{Horak2009} P. Horak et al., arXiv:0904.3059, A. Xuereb et al., Phys. Rev. A \textbf{79}, 053810 (2009)
     %A. Xuereb et al., Phys. Rev. A \textbf{80}, 013836 (2009)



\bibitem{Shampire95} P. Samphire et al., Phys. Rev. A.
\textbf{51}, 2726 (1995).

\bibitem{Gardiner1993} C.W. Gardiner, Phys. Rev. Lett. \textbf{70}, 2269 (1993), H.J. Carmichael, Phys. Rev. Lett. \textbf{70}, 2273 (1993)

\bibitem{Hammerer2009} K. Hammerer, A.S. Sorensen, E.S. Polzik, to be published in Rev. Mod. Phys., arXiv:0807.3358

\end{thebibliography}
\end{document}